\documentclass{INTERSPEECH2023}

\usepackage{graphicx, float,verbatim}
\usepackage{algorithmic, amsfonts, gensymb}
\usepackage{amsmath, amssymb}
\usepackage{cite, array, multirow}
\usepackage{setspace,bbding, hyperref}
\interspeechcameraready

\title{3D Neural Beamforming for Multi-channel Speech Separation \\ Against Location Uncertainty}
\name{Rongzhi Gu$^1$, Shi-Xiong Zhang$^2$, Dong Yu$^2$}
\address{
  $^1$ Tencent AI Lab, Shenzhen, China\\
  $^2$ Tencent AI Lab, Seattle, WA, USA
}
\email{\{lorrygu,auszhang,dyu\}@tencent.com}

\begin{document}

\maketitle

\begin{abstract}
Multi-channel speech separation using speaker's directional information has demonstrated significant gains over blind speech separation. However, it has two limitations. First, substantial performance degradation is observed when the coming directions of two sounds are close. Second, the result highly relies on the precise estimation of the speaker’s direction.
To overcome these issues, this paper proposes 3D features and an associated 3D neural beamformer for multi-channel speech separation. Previous works in this area are extended in two important directions. First, the traditional 1D directional beam patterns are generalized to 3D. This enables the model to extract speech from any target region in the 3D space. Thus, speakers with similar directions but different elevations or distances become separable. Second, to handle the speaker location uncertainty, previously proposed \textit{spatial} feature is extended to a new 3D \textit{region} feature.
The proposed 3D region feature and 3D neural beamformer are evaluated under an in-car scenario. Experimental results demonstrated that the combination of 3D feature and 3D beamformer can achieve comparable performance to the separation model with ground truth speaker location as input.

\end{abstract}
\noindent\textbf{Index Terms}: 3D beam pattern, neural beamformer, 3D feature, speech separation, location uncertainty

\section{Introduction}
\label{sec:intro}
Target speech separation (TSS) aims to recover the target speech from reverberant noisy mixture, which is one of the most important yet challenging tasks for robust automatic speech recognition \cite{barker2018fifth,yoshioka2018multi,wang2018supervised}.
With the advance of deep learning, one of the most successful multichannel TSS (MC-TSS) schemes is time-frequency (T-F) masking-based beamforming, which combines the deep neural network (DNN) with well-designed beamforming techniques \cite{du2016ustc,heymann2016neural,erdogan2016improved,higuchi2016robust}. The DNN is trained to estimate a T-F mask to calculate the signal statistics, such as spatial covariance matrices (SCMs), for the subsequent beamforming algorithm. Recently, the two-stage T-F masking-based beamforming scheme is transferred to a neural beamforming scheme \cite{heymann2017beamnet, ochiai2017unified}, where the beamforming algorithms are implemented as fully differentiable network layers and operations for end-to-end training. Very recently, Xu et. al. \cite{xu2021generalized,zhang2020adl} propose an all-neural beamforming (AN-BF) framework, which integrates mask estimation and parametric beamforming coefficient estimation into a unified network that can be trained from end-to-end.

To train a more powerful mask estimation network, direction-aware MC-TSS methods assume that the direction-of-arrival (DOA) of the target speaker is pre-estimated or pre-defined by the usage, and develop different kinds of spatial features \cite{wang2018spatial,chen2018multi,gu2019neural} to explicitly indicate the dominance of directional sources for better mask estimation. Despite the impressive improvements achieved by direction-aware MC-TSS methods, the strong dependency on precise DOA estimation is not trivial. The DOA estimation error severely deteriorates separation performance especially when the azimuths of simultaneous speech are close \cite{gu2020multi}. Although the DOA mismatch issue has been extensively investigated for robust beamforming \cite{li2003robust,khabbazibasmenj2012robust,yu2006robust,zhang2013robust}, the solutions for neural network based beamforming has not been fully explored.

To release the burden of precise source localization while accounting for the location uncertainty, this work proposes a 3D AN-BF method for MC-TSS. Firstly, the target speaker is assumed to locate within a limited 3D region centered at the estimated location. The 3D setup is adopted to enable the MC-TSS model to distinguish sources with close azimuths via their different elevations and source-to-array distances. Then, a 3D region feature is designed condition on the vertices and center of the region. Via a learning based attention module, the 3D region feature learns to aggregate and attend to different spatial views of the region. The 3D region feature is served as the input to an AN-BF network to advance beamforming weights estimation.
To evaluate our proposed method, we consider an in-car scenario, where potential speakers are located within fixed regions (seats). With only the center location of the target seat provided, and an approximate region boundary estimated, the proposed method exhibits comparable target separation performance to models with ground truth location as input.

The rest of the paper is organized as follows. Section \ref{sec:3d_rf} reviews the 3D spatial feature and proposes a 3D region feature. Section \ref{sec:anb} describes the proposed 3D AN-BF method. Section \ref{sec:exp} presents the experiments and analyzes the results. Section \ref{sec:conclusion} concludes the paper.

\section{3D Feature}
\label{sec:3d_rf}
\subsection{3D spatial feature}
\label{subsec:3d_sf}

\begin{figure}
\centerline{\includegraphics[width=\linewidth]{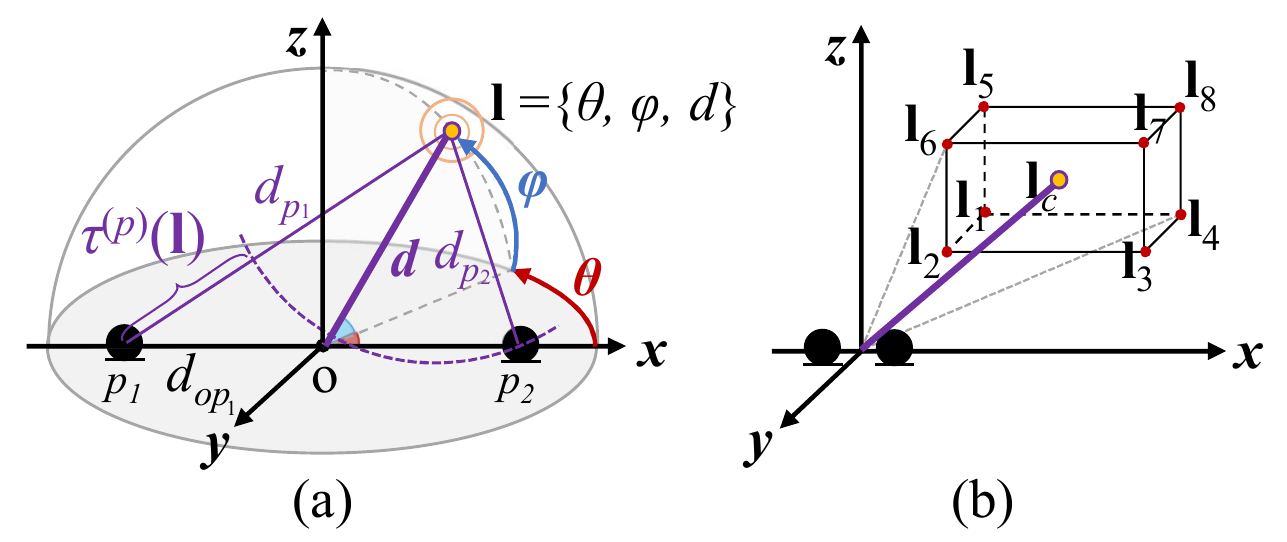}}
%\caption{Illustrations of (a) 3D spatial feature formulation; (b) 3D candidate region; (c) azimuth and elevation grid of 3D sphere feature. The yellow point denotes the given location $\mathbf{l}_c$ as the center of the region, and $\mathbf{l}_i$ marks the potential vertices or sampling location of a 3D region.}
\caption{Illustrations of (a) 3D spatial feature formulation; (b) 3D candidate region. The yellow point denotes the given location $\mathbf{l}_c$ as the center of the region, and $\mathbf{l}_i$ marks the potential vertices or sampling location of a 3D region.}
\label{fig:3d}
\end{figure}

To alleviate the spatial ambiguity issue when simultaneous speech come from close azimuths, a 3D spatial feature was proposed in our recent work \cite{gu20213d}. Specially designed for near-field speech applications, the 3D spatial feature is developed based on the spherical wave propagation model and assumes the availability of the location information $\mathbf{l}=\{\theta, \phi, d\}$ including azimuth $\theta$, elevation $\phi$ and distance from the target source to the array center $d$. The 3D information empowers the 3D spatial feature to more precisely indicate the dominance degree of the target speech at each T-F bin. As shown in Figure \ref{fig:3d} (a), the pure delay $\tau^{(p)}(\mathbf{l})$ between the $p$-th microphone pair is computed as the delay between two direct source-to-microphone paths, $d_{p_1}$ and $d_{p_2}$:
\vspace{-0.1cm}
\begin{equation}
\tau^{(p)}(\mathbf{l})=
 (d_{p_1}-d_{p_2})f_s / c\\
\label{eq:tf_3dtau}
\end{equation}
where $c$ is the sound velocity, $f_s$ is the sampling rate, $d_{p_1}$ and $d_{p_2}$ are the distances between the target speaker and the $p_1$-th and the $p_2$-th microphone, respectively. According to the law of cosines, $d^2_{p_1}$ can be computed with $d^2_{p_1}=d^2_{op_1}+d^2-2d_{op_1}d \cos \alpha$, where $\cos\alpha=\cos\theta\cos\phi$, $d_{op_1}$ is the distance between microphone center $o$ and microphone $p_1$. Similarly, $d^2_{m_2}$ can be computed. The 3D spatial feature is derived by comparing the similarity between the observed interaural phase differences (IPDs) and theoretical interaural phase differences (TPD), and the match degree will indicate the dominance of the source at location $\mathbf{l}$ at each T-F bin \cite{chen2018multi}:
\begin{equation}
\textit{SF}_{t,f}(\mathbf{l})={\sum}_p\left <
\text{IPD}^{(p)}_{t,f},\text{TPD}^{(p)}_f(\mathbf{l})
\right >
\label{eq:tf_df}
\end{equation}
where $\text{IPD}^{(p)}_{t,f}=\angle\mathbf{Y}^{(p_1)}_{t,f}-\angle\mathbf{Y}^{(p_2)}_{t,f}$, $\text{TPD}^{(p)}_{f}(\mathbf{l})=2\pi f\tau^{(p)}(\mathbf{l})$, $\mathbf{Y}$ is the multichannel complex spectrogram and $t$, $f$ respectively index the frame and frequency band.

\subsection{3D region feature}
\label{subsec:3d_rf}

However, the 3D spatial feature is sensitive to the location estimation error, which brings about extra burden for precise sound localization.
To account for the uncertainty of the location information, i.e., inaccurate source localization, array and camera miscalibration, this work makes an attempt to learn a robust model by posing the potential location deviations at the training stage.

A straightforward method is to introduce random perturbations to the given azimuths, elevations and distances as the new input to the model. This method may mislead the model to learn a broader main beam to tolerate the errors, therefore degrading the separation performance.

To fully unleash the power of the 3D setup and 3D spatial feature, this work proposes a 3D region feature. Motivated by Bayesian beamforming \cite{lam2006bayesian,bell2000bayesian}, we assume each source is located within a limited 3D region (3D box in this work, without loss of generality), the center of which is the estimated location of the target source.
As illustrated in Figure \ref{fig:3d} (b), with the availability of elevation and distance, except for the given location, extra vertices (e.g., $\mathbf{l}_1$ to $\mathbf{l}_8$) of the 3D region can be sampled to take a full spatial view of the whole region. The desired region feature is modeled as the mixture of candidate 3D spatial features combined with the posterior distribution of the candidate locations $\mathbf{l}_i$:
\begin{equation}
\textit{RF}(\mathbf{l})=\sum_{i=1}^{\mathcal{L}} p \left(\mathbf{l}_i | \textit{SF}(\mathbf{l}_1),...,\textit{SF}(\mathbf{l}_\mathcal{L}) \right) \textit{SF}(\mathbf{l}_i)
\label{eq:combine_sf}
\end{equation}
where Eq. \ref{eq:combine_sf} omits $t,f$ index for simplification,
$\textit{SF}(\mathbf{l}_i)$ is computed using location information of vertex $\mathbf{l}_i$, $\mathcal{L}$ is the total number of vertices including the center,  %$\mathbf{L}=\{\mathbf{l}_c, \mathbf{l}_1,...,\mathbf{l}_C\}$ is the set of all vertices,
$p \left(\mathbf{l}_i | \textit{SF}(\mathbf{l}_1),...,\textit{SF}(\mathbf{l}_\mathcal{L}) \right)$ is the posterior of the source existence at $\mathbf{l}_i$, estimated via an attention module optimized with the separation network:
\begin{equation}
p \left(\mathbf{l}_i | \textit{SF}(\mathbf{l}_1),...,\textit{SF}(\mathbf{l}_\mathcal{L}) \right)
= \sum_{t} {g\left([\textit{SF}_t(\mathbf{l}_1),...,\textit{SF}_t(\mathbf{l}_C)]^\mathsf{T}\right)}
\label{eq:att}
\end{equation}
where $g(\cdot)$ denotes two fully connected (FC) layers to estimate the posterior distribution of spatial features computed with $\mathcal{L}$ location candidates. Instead of combining the directional beamformer candidates at the output side, we try to refine the 3D region feature before the mask estimator to save computational cost.

\section{3D All-Neural Beamforming}
\label{sec:anb}

\begin{figure}
\centerline{\includegraphics[width=\linewidth]{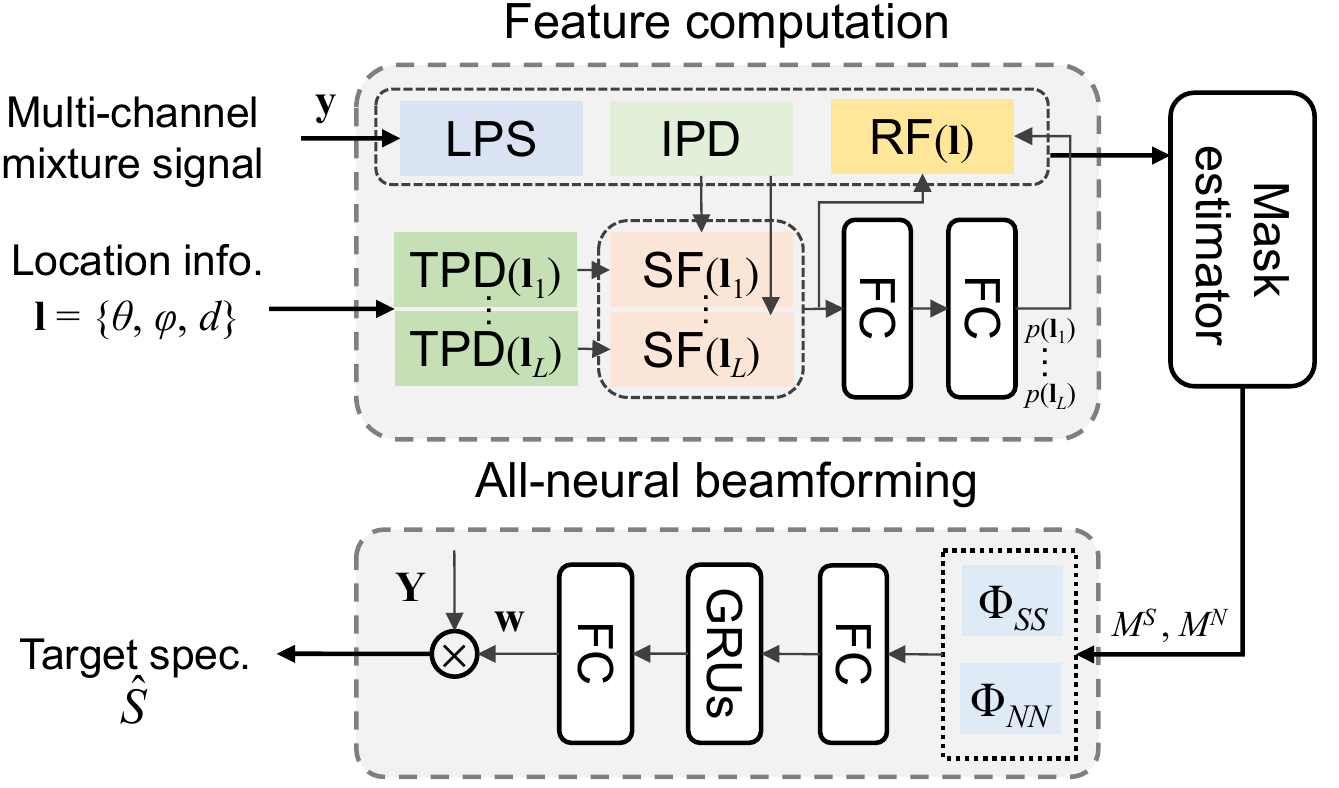}}
\caption{The framework of the proposed 3D all-neural beamforming method, where $\mathbf{y}$ is the multi-channel mixture, $\hat{s}$ is the estimated target speech.}
\label{fig:framework}
\end{figure}

The overall framework of the 3D all-neural beamforming is illustrated in Figure \ref{fig:framework}, aiming to separate the target speech $s$ from the multichannel mixture $\mathbf{y}$, given the coarse location information $\mathbf{l}$ of the target speaker. This kind of location information can be obtained via a depth camera, pre-defined in real usage, or pre-estimated by a sound localization frontend.

The proposed framework consists of 3 modules: 3D feature computation, mask estimation and all-neural beamforming. Following our previous work \cite{gu20213d}, the combination of logarithm power spectra (LPS), IPD and 3D region feature formed by the target location information is adopted as the input feature. The features are then fed into a deep neural network based mask estimator to estimate the T-F masks of the target speech $M^S$ and interfering speech $M^N$. At the AN-BF stage \cite{xu2021generalized}, the estimated mask is utilized to compute the time-varying spatial covariance matrices (SCMs) of the target and interfering speech:
%To remove the burden of global normalization and mask estimation for interfering speech, the proposed method computes the following spatial covariance matrices:
\begin{equation}
\begin{split}
\Phi^\textit{SS}_{t,f} &= \mathbf{\hat{S}}_{t,f} \mathbf{\hat{S}}^{\mathsf{H}}_{t,f}\\
\Phi^\textit{NN}_{t,f} &= \mathbf{\hat{N}}_{t,f} \mathbf{\hat{N}}^{\mathsf{H}}_{t,f}
\end{split}
\label{eq:fd_scm}
\end{equation}
where $\mathbf{\hat{S}}_{t,f}=M^S_{t,f} \circ \mathbf{Y}_{t,f}$, $\mathbf{\hat{N}}_{t,f}=M^N_{t,f} \circ \mathbf{Y}_{t,f}$ are the computed target and interfering spectrogram based on estimated masks, $(\cdot)^\mathsf{H}$ denotes the complex conjugate matrix. These two SCMs are served as the input of the AN-BF network, which is consist of FC layers and gated recurrent unit (GRU) layers to estimate the time-varying beamforming coefficients $\mathbf{w}$. The final target estimation $\hat{S}$ is then obtained by applying $\mathbf{w}$ to the multichannel spectrograms:
\begin{equation}
%\frac{\left ( M_S \circ \mathbf{Y} \right ) \left ( M_S \odot \mathbf{Y} \right ) ^\mathsf{H}}{M_S M_S^\mathsf{H}}
\hat{S}_{t,f} = \mathbf{w}^{\mathsf{H}}_{t,f}\mathbf{Y}_{t,f}
\label{eq:fd_bf}
\end{equation}
The final estimation is then converted back to waveform $\hat{s}$ using inverse STFT. The whole framework is trained from end-to-end using scale-invariant signal-to-distortion ratio (SI-SDR) as the loss function \cite{le2019sdr}.

\section{Experimental Setup}
\label{sec:exp}
\subsection{Data preparation}
\label{subsec:data}
\begin{figure}[h]
% \centerline{\includegraphics[width=4cm]{figs/car_4region.pdf}}
\centerline{\includegraphics[width=\linewidth]{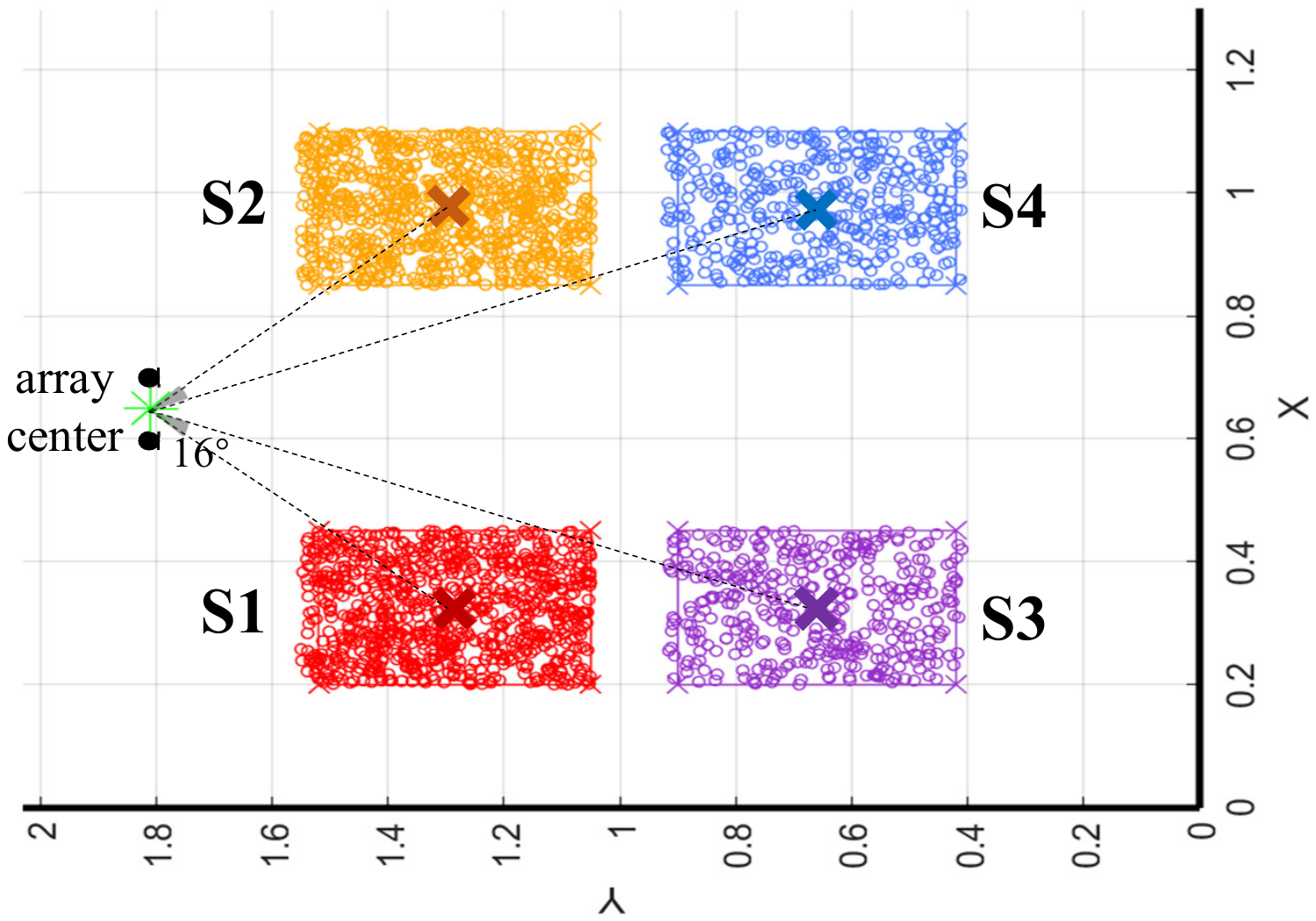}}
\caption{The top view of an in-car scenario with 4 potential speakers. Each point indicates a potential source location and the crosses mark the centers of the regions. The azimuth difference between the centers of S1 and S3 is 16$\degree$. }
\label{fig:car_region}
\end{figure}

We consider an in-car scenario to evaluate the proposed method in real-world applications. As shown in Figure \ref{fig:car_region}, there are 4 potential speakers and their corresponding regions in a car: the main driver (S1), the co-driver (S2) and two passengers (S3 \& S4) sitting in the back. The main driver's voice is taken as the target.  It can be seen from the top view that the azimuths of the main driver (S1) and the passenger in the back seat (S3) are very close. In this case, it is difficult to distinguish these two speakers with the spatial feature only based on azimuth.

The data is simulated using AISHELL-2 corpus, containing 90 k, 9 k and 2 k cochannel noisy reverberant mixtures for training, validation and evaluation, respectively. There are up to 3 speakers in the mixture and the main driver is always speaking. The multi-channel signals are generated using image-source method (ISM) [15]. We use a dual mic with 11.8 cm spacing. The reverberation time T60 is ranging 0.05s to 0.7 seconds. The room size matches that of the car and the microphone array is placed at the car head. The signal-to-interference ratio (SIR) is ranging from -6 to 6 dB. Also, we add at least 3 directional noises with signal-to-noise ratio (SNR) ranging from -5 to 20 dB. All data is sampled at 16k Hz.

To simulate the potential sources' locations in different regions, we refer to the situation of the in-car scenario. The sitting height is set in the range of $[0.95, 1.15]$ m. The 3D box boundary is decided according to the head size (about 0.2 m) and the seat width of the car.

\subsection{Features, Network and Training details}
For short time Fourier transform setting, we use 32ms square-root Hann window and 16ms hop size, resulting in $F=257$ frequency bands. Only one microphone pair (1,2) is adopted. The input feature dimension is $T\times 3F$.

Conv-TasNet \cite{luo2019convtasnet} is served as the mask estimator. For the attention module, the first FC layer receives the feature size of $\mathcal{L}F$ and transforms into $40$. The second FC layer outputs $\mathcal{L}$ probabilities for each candidate location. For all-neural beamforming network, the first FC layer that processes the computed SCMs has 180 cells and the hidden size of two GRU layers is 180 and 90, respectively. The output size of the final FC layer is $2F$.

All the models are trained with 4-second mixture chunks, using Adam optimizer with early stopping. Learning rate is initialized as 1e-3 and will be decayed by 0.5 when the validation loss has no improvement for consecutive 3 epochs.

\subsection{Evaluation setup}

SI-SDR, perceptual evaluation speech quality (PESQ) and Chinese character error rate (CER) are adopted as the evaluation metrics and the reverberant clean target speech is used as reference for all the metric computations.% to measure the separation quality and speech intelligibility. %The CER is measured using Tencent commercial mandarin speech recognition API \cite{tencentapi}.

The performance is evaluated under different speaker mixing conditions: 1 speaker (only S1), 2 speakers (S1+3 is harder since the azimuths are close), and 3 speakers. We consider 4 kinds of training and evaluation cases: 1) GT-GT: Use ground truth (GT) azimuth or location as input at both stages; 2) GT-fix: Train with GT while testing with fixed region center (fixed $\theta_c$ or $\mathbf{l}_c$); 3) fix-fix: Use fixed region center as input at both stages; 4) Use candidate vertices as well as the region center ($\mathcal{L}$=8+1) as input at both stages.

For reference, we also compute some oracle beamforming results, including oracle minimum variance distortionless response (MVDR) computed with ideal ratio masks, oracle multichannel Wiener filter (MCWF) computed with GT target spectrogram.

\begin{table*}[hbt]
  \caption{SI-SDR (dB), PESQ and CER (\%) results of different input feature and output target configurations of MC-TSS models. }
  \label{tab:rlt}

  \centering
  \scalebox{0.92}{
  %\begin{tabular}{l|c|c|c|p{20 pt}<{\centering}p{20 pt}<{\centering}p{20 pt}<{\centering}|p{20 pt}<{\centering}|c}
  \begin{tabular}{l|c|c|c|ccc|ccc|c|c|c}
    \hline
     \hline
    \multirow{2}{*}{\textbf{Approach}} &
    \multirow{2}{*}{\textbf{Train}} &
    \multirow{2}{*}{\textbf{Eval}} &
    \multicolumn{8}{c|}{\textbf{SI-SDR (dB)} $\uparrow$} &
    \multirow{2}{*}{\textbf{PESQ} $\uparrow$} &
    \multirow{2}{*}{\textbf{CER}(\%) $\downarrow$} \\
    % \multirow{2}{*}{\textbf{RTF}}\\
    & & & S1 & S1+2 & S1+3 & S1+4 & S1+2+3 & S1+2+4 & S1+3+4 & Ave. & & \\
    \hline
    Mixture & - & - & 4.18 & -4.97 & -5.21 & -5.31 & -7.27 & -6.64 & -6.86 & -4.99 & 1.81 & 91.40 \\
    % Reverb. clean  & - & - & $\infty$ & $\infty$ &$\infty$ & $\infty$ & 4.50 & ? \\
    \hline
     \hline
    1D-cRM &  GT & GT & 15.72 & 8.80 & 7.36 & 8.91 & 6.10 & 6.67 & 6.69 & 8.24 & 2.28 & 28.57 \\ % 29.90
    1D-cRM &  GT  & fixed $\theta_c$  & 15.80 & 8.17 & 7.13 & 8.62 & 5.57 & 4.81 & 4.27 & 7.12 & 2.21 & 30.91 \\
    1D-cRM &  fixed $\theta_c$ & fixed $\theta_c$  & 14.01 & 6.72 & 7.01 & 7.32 & 2.81 & 4.33 & 6.03 & 5.87 & 2.04 & 58.17 \\
    \hline
    3D-cRM & GT & GT & 16.50 & 8.94 & 8.78 & 9.11 & 6.93 & 7.30 & 7.13 & 8.74 & 2.34 & 24.68 \\
    3D-cRM & GT & fixed $\mathbf{l}_c$ & 6.73 & 3.47 & -1.00 & 4.33 & 0.54 & -0.25 & 2.03 & 0.80 & 2.10 & 57.72 \\
    3D-cRM & fixed $\mathbf{l}_c$ & fixed $\mathbf{l}_c$ & 13.91 & 7.97 & 7.82 & 8.46 & 3.33 & 4.85 & 6.31 & 6.57 & 2.11 & 42.54 \\
    3D-cRM & $\{\mathbf{l}_i\}_{i=1}^{9}$ & $\{\mathbf{l}_i\}_{i=1}^{9}$ & 15.99 & 8.36 & 7.83 & 8.45 & 6.07 & 6.21 & 6.35 & 7.94 & 2.23 &  29.28 \\
    \hline
    \hline
    1D-AN-BF  & GT & GT & 18.25 & 10.07 & 9.17 & 10.21 & 7.33 & 7.87 & 7.62 & 9.45 & 2.72  & 13.72 \\
    1D-AN-BF  & GT & fixed $\theta_c$ & 18.21 & 9.48 & 7.83 & 9.79 & 6.50 & 7.01 & 7.05 & 8.63 & 2.67 & 20.80 \\
    1D-AN-BF  & fixed $\theta_c$ & fixed $\theta_c$ & 18.04 & 10.01 & 9.12 & 10.15 & 7.16 & 7.85 & 7.55 & 9.36 & 2.72 & 14.68 \\ % 这个结果需要重新看一下
    \hline
    3D-AN-BF  & GT & GT & 18.68 & 10.33 & 9.77 & 10.44 & 7.48 & 8.27 & 7.94 & 9.77 & 2.76 & \textbf{12.31} \\
    3D-AN-BF  & GT & fixed $\mathbf{l}_c$ & 13.13 & 5.94 & 3.56 & 6.81 & 3.56 & 2.90 & 4.31 & 4.78 & 2.45 & 31.59 \\
    3D-AN-BF  & fixed $\mathbf{l}_c$ & fixed $\mathbf{l}_c$ & 18.57 & 10.41 & 9.23 & 10.45 & 7.41 & 7.88 & 7.83 & 9.64 & 2.75 & 15.28 \\
     % 3D-AN-BF  & $\{\mathbf{l}_i\}_{i=1}^{5}$ & $\{\mathbf{l}_i\}_{i=1}^{5}$ & 23.4 & 10.6 & 9.0 & 10.4 & 28.56 \\
     3D-AN-BF  & $\{\mathbf{l}_i\}_{i=1}^{9}$ & $\{\mathbf{l}_i\}_{i=1}^{9}$ & 18.90 & 10.87 & 9.79 & 10.77 & 7.64 & 8.31 & 8.13 & \textbf{10.01} & \textbf{2.80} & 12.97 \\

    \hline
 \hline
  oracle MVDR & - & - & 7.44 & 0.94 & -0.15 & 0.46 & -1.67 & -1.52 & -1.74 & 0.18 & 2.18 & 69.02 \\
  oracle MCWF & - & - & 15.78 & 6.08 & 5.11 & 5.98 & 3.31 & 3.49 & 3.43 & 5.52 & 2.53 & 55.85 \\
%  super-directive BF & - & fixed $\theta_c$ & \\
%  delay-and-sum BF & - & fixed $\theta_c$ & \\
 \hline
 \hline
  \end{tabular}}
\end{table*}

\section{Results analysis}
\label{sec:result}

\subsection{3D beampattern visualization}
\begin{figure}[t]
\centerline{\includegraphics[width=\linewidth]{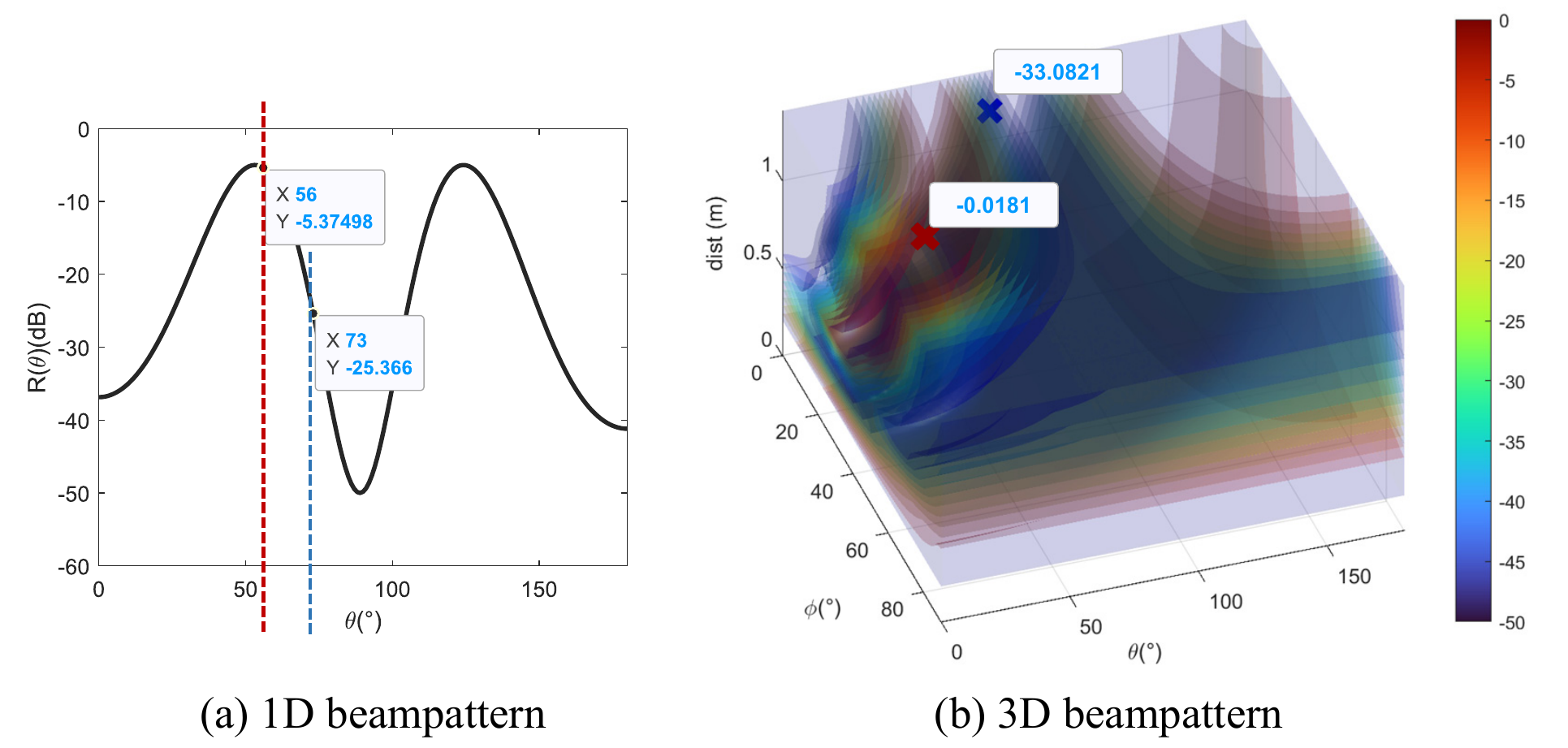}}
\caption{An example of 1D and 3D beampatterns derived from 1D-AN-BF and 3D-AN-BF methods, respectively. The azimuths of the target source (red cross in region S1) and interference source (blue cross in region S3) are $56.3\degree$ ad $72.5\degree$, respectively.}
\label{fig:3d_bfp}
\end{figure}

Figure \ref{fig:3d_bfp} visualizes an example of 1D and 3D beamforming patterns of the corresponding beamformers. The $x, y, z$ axis respectively represents azimuth $\theta$, elevation $\phi$ and distance $d$. The azimuth difference between the target and interfering source is rather small, i.e., $16.2\degree$.

It can be observed that the main beam in (b) is steered to the target azimuth at the corresponding elevation and distance grid while suppressing the interfering speech, which exhibits a more desiring property than the 1D pattern.

\subsection{Performance evaluation}
Table \ref{tab:rlt} reports the SI-SDR, PESQ and CER results of different training and evaluation setups.

Firstly, comparing the S1+3 results of 1D and 3D-cRM models with GT-GT inputs, i.e., 7.36 v.s. 8.78 dB, the effectiveness of 3D spatial feature for alleviating spatial ambiguity issue is demonstrated. However, when encountering estimation errors (GT-fix), the 3D spatial feature is more sensitive and the model performance degrades severely. Considering the estimation errors during training (fix-fix) can be helpful, yet not satisfactory.

Then, for AN-BF models, the AN-BF stage markedly improves the PESQ and CER metrics and slightly alleviates the sensitivity to the estimation error.

Equipped with the proposed 3D region feature, the 3D-AN-BF ($\{\mathbf{l}_i\}_{i=1}^{9}$-$\{\mathbf{l}_i\}_{i=1}^{9}$) model achieves comparable performance with GT-GT setup. In this way, the need for precise localization is mitigated.

\subsection{Evaluation on real-recorded data}
We evaluate the proposed method with different input features and output targets on 25-min real data recorded in a driving car, where speech signals picked from AISHELL are replayed according to pre-arranged timestamps in each region.

All the models are trained on simulated data in Section \ref{subsec:data}. The results are reported in Table \ref{tab:real_rlt}, where the CER is measured using the AISHELL transcript. Although there is mismatch (e.g., echo, RIR, music) between the training data and real-recorded data, the recognition results are consistent with those on simulation data. Compared to the 3D-cRM with CER of 42.7\%, the proposed 3D neural beamforming method decreases the CER by 35.8\%. With the region feature aggregation, the CER is further reduced by 6.9\%.

For more audio samples and details, please refer to \footnote{\href{https://moplast.github.io/3d.github.io}{https://moplast.github.io/3d.github.io}}.

\begin{table}[h]
    \caption{CER (\%) results on real-recorded data.}
    \label{tab:real_rlt}
    \centering
\scalebox{0.92}{
    \begin{tabular}{c|ccccc}
    \hline\hline
    \textbf{Feature} & \textbf{Mix.} & \textbf{1D} ($\theta_c$) & \textbf{3D} ($\mathbf{l}_c$) &\textbf{3D} ($\mathbf{l}_c$) &\textbf{3D} ($\{\mathbf{l}_c\}^9$) \\
    \textbf{Target} & - & \textbf{cRM} & \textbf{cRM} & \textbf{AN-BF} & \textbf{AN-BF} \\
    \hline
    CER (\%) &  108.7 & 45.9 & 42.7 & 27.4 & 25.5 \\
    \hline\hline
    \end{tabular}
    }
\end{table}

\section{Conclusion}
%\vspace{-0.3cm}
\label{sec:conclusion}
This work proposed a 3D neural beamforming method for multi-channel speech separation to release the burden of precise source localization while accounting for the location uncertainty. A 3D region feature was designed to extract and selectively attend to different spatial views within a candidate region, and then integrated into an all-neural beamforming network. The evaluation results under an in-car scene, on both simulated data and real-recorded data, demonstrated the effectiveness of the proposed method.

\bibliographystyle{IEEEtran}
\bibliography{main}

\end{document}